\documentclass{article}
\usepackage{spconf,amsmath,graphicx,cite,amssymb,url}
\usepackage{graphicx}
\usepackage{subcaption}
\usepackage{color}

\def\WASCHL{WASCHL}

\title{Sparse DOA Estimation of Wideband Sound Sources Using Circular Harmonics}
\name{Clemens Hage\thanks{This work has been supported by the German BMBF under the funding code 16SV5803, and by the DFG cluster of excellence CoTeSys}, Tim Habigt, Martin Kleinsteuber}
\address{Technische Universit\"{a}t M\"{u}nchen\\Fakult\"{a}t f\"{u}r Elektro- und Informationstechnik\\Arcisstr. 31, D-80333 M\"{u}nchen}
\begin{document}

\maketitle
\begin{abstract}
Sparse signal models are in the focus of recent developments in narrowband DOA estimation. Applying these methods to localizing audio sources, however, is challenging due to the wideband nature of the signals. The common approach of processing all frequency bands separately and fusing the results is costly and can introduce errors in the solution. We show how these problems can be overcome by decomposing the wavefield of a circular microphone array and using circular harmonic coefficients instead of time-frequency data for sparse DOA estimation. As a result, we present the super-resolution localization method \WASCHL{} (Wideband Audio Sparse Circular Harmonics Localizer) that is inherently frequency-coherent and highly efficient from a computational point of view. 
\end{abstract}

\section{Introduction}
Direction of arrival (DOA) estimation aims at retrieving the position of multiple active signals in a scene. Considering $K$ narrowband sources of center frequency $\omega_0$, the commonly used signal model in the time-frequency domain writes as
\begin{equation}
\label{eq:tfmodel}
\mathbf{y}_{n}(\omega_0) = A(\omega_0) \mathbf{x}_{n}(\omega_0) + \mathbf{w}_{n},
\end{equation}
where $\mathbf{y}_n$ contains the Short Time Fourier Transform (STFT) of $M$ microphone inputs at time instance $n$, $\mathbf{w}_n$ is additive Gaussian noise, and $A(\omega_0)$ contains information about the delays between the microphone inputs for the respective $K$ sources. Considering a planar circular array of equispaced omnidirectional sensors placed at a radius $R$ and assuming all sources are in the same plane as the array, the columns of $A(\omega_0) \in \mathbb{C}^{M \times K}$ are
\begin{equation}
\label{eq:arraymatrix}
\mathbf{a}_{i}(\omega_0) = \begin{bmatrix} e^{-j \tfrac{\omega_0}{c}R \cos(\theta_{i} - \theta_1)} & \dots & e^{-j \tfrac{\omega_0}{c}R \cos(\theta_{i} - \theta_M)} \end{bmatrix}^\top
\end{equation}
with $\theta_{i}, i = 1, \dots, K$ being the azimuth angles defining the original incident direction of the true sources, $\theta_m$ defining the positions of the $m=1,\dots,M$ sensors and $j=\sqrt{-1}$.

A common way to estimate the DOA of the signals from the input is to define a search grid of locations $\theta_q, q=1,\dots,Q$ with $Q \gg K$ around the array and to determine the incident direction of the original sources by evaluating a cost function for all these positions. In the course of investigating sparse signal models a method has been proposed in \cite{Malioutov2005} that casts DOA estimation as a sparse coding problem. In this approach an analytic dictionary $D(\omega_0) \in \mathbb{C}^{M \times Q}$ is introduced, whose columns have the same structure as \eqref{eq:arraymatrix}, the difference being that the columns stem from the predefined grid locations $\theta_q$ instead of the unknown $\theta_i$. This motivates the data model
\begin{equation}
\label{eq:L1SVD}
\mathbf{y}_n(\omega_0) = D(\omega_0)\mathbf{s}_n(\omega_0) + \mathbf{w}_n.
\end{equation}
Assuming ideal conditions, $\mathbf{s}_n(\omega_0)$ is a sparse vector and its support is identical to the positions of the active sources. In \cite{Malioutov2005} it has been shown that this data model leads to a convex sparse coding approach for DOA estimation of narrowband sources, which is commonly referred to as the L1-SVD method. Especially, the method is robust against overestimating the total number of sources in the scene, which is one of the limitations of the well-known MUSIC approach \cite{Schmidt1986}. The main drawback, on the other hand, is that the delays at the microphones and thus the dictionary $D(\omega)$ are frequency-dependent. This becomes a serious obstacle when we turn from narrowband to wideband signals, such as audio sources, which occupy a wide part of the spectrum.

The authors of \cite{Malioutov2005} discuss three general ideas to make their approach applicable to wideband signals. An intuitive approach is to estimate $\mathbf{s}_n(\omega)$ for all frequencies separately and to average over the results. But due to the separate processing the estimation of $\mathbf{s}_n$ is incoherent as its support can vary over the frequency range. The second option, which is concatenating all measurements and dictionaries to solve a joint problem, is not viable in practice as the dimension becomes intangibly large already with a few simultaneously considered frequencies. The third approach introduces wideband focusing matrices that align the steering vectors for all frequency bins in a frequency region of interest. However, this processing needs an initial coarse estimate of all source positions, and the approximation quality deteriorates with the distance in frequency \cite{Malioutov2005}. Recently, Luo et al.~\cite{Luo2013} have proposed a subband information fusion method that extends the separate-frequencies approach by an additional penalty term to force coherence between all frequency bands. Instead of encoding the inputs separately, an overcomplete dictionary is computed out of all the individual inputs and one overall optimization problem is solved to obtain a so-called sparse indicative vector related to the estimated incident direction. On the upside the frequency-coherence is improved as simulated experiments indicate, but on the downside a new dictionary needs to be computed for each input.

A promising approach of inherently frequency-coherent DOA estimation is to decompose the wavefield and to transfer the input from the time domain to the modal domain, cf. \cite{Teutsch2006}. In \cite{Torres2012} it has been shown how the DOA of several speech signals can be estimated by evaluating the circular harmonic (CH) coefficients instead of the time-frequency data. The proposed approach transfers the concept of the well-known delay-and-sum beamformer (DSB) to the circular harmonics beamformer (CHB), as the STFT coefficients of the input signals are transformed into the modal domain through applying a spatial Fast-Fourier-Transform (FFT). By equalizing the CH coefficients at each frequency an array matrix can be used that is frequency-independent, i.e. in contrast to \eqref{eq:arraymatrix} the steering vectors do not change with the frequency.

The contribution of this work is to introduce the circular harmonics concept to sparsity-based DOA estimation of wideband sound signals using a circular microphone array. We propose a method for localizing an unknown number of wideband sound sources that is coherent over the frequency range and is computationally cheap. In the following we will recall the required concepts and show how to perform sparsity-based DOA estimation with a frequency-independent dictionary using circular harmonics. Simulations and experiments on recorded data compare the approach with the frequency-separated approach and a circular harmonics beamforming technique and show that the proposed algorithm is a viable approach for computationally light DOA estimation of multiple wideband sound sources.

\section{Circular harmonic analysis}
Wavefield analysis for circular microphone arrays has been studied in the past years \cite{Teutsch2006,Tiana2010,Parthy2011}. The general concept is to assume a continuous circular aperture and to decompose the distribution of sound pressure at each point into a series of eigensolutions of the acoustic wave equation.  According to \cite{Teutsch2006}, the momentary pressure at a point $(\theta,R)$ and caused by a plane wave front coming from the direction $\theta_i$ can be written as
\begin{equation}
\label{eq:idealdecomp}
p(k,R,\theta) = p_0 \sum_{p=-\infty}^\infty c_p (k,R, \theta_i) e^{jp\theta}
\end{equation}
for a certain wave number $k = \tfrac{\omega}{c}$ depending on the observed frequency $\omega$.
The coefficients $c_p (k,R, \theta_q)$ in this Fourier series are
\begin{equation}
\label{eq:idealcoeffs}
c_p (k,R,\theta_q) = p_0 j^p J_p(kR)e^{-jp \theta_q},
\end{equation}
with $p_0$ being the amplitude of the sound wave and $J_p(kR)$ the Bessel function of the first kind of order $p$ evaluated at $kR$.

Clearly, this ideal decomposition of the wave field cannot be performed in practice as a real microphone array is not a continuous aperture but consists of a finite number of discretely placed sensors. As a consequence, the maximum order $L$ of modes that can be observed is limited to $\lfloor \tfrac{M-1}{2} \rfloor$, and the infinite summation in \eqref{eq:idealdecomp} becomes a finite sum over $p=-L,\dots,L$.
The approximated inverse transform to \eqref{eq:idealdecomp} that computes the CH coefficients from the microphone inputs up to a certain error \cite{Teutsch2006,Torres2012} writes as
\begin{equation}
\label{eq:sampledcoeffs}
\tilde{c}_p(k,R,\theta_q) = \tfrac{1}{M} \sum_{m=0}^{M-1} \tilde{p}_m(k,R,\theta_q) e^{-jp\theta_m}
\end{equation}
where $\tilde{p}_m(k,R,\theta_q)$ is the detected excitation at the $m$-th microphone caused by a sound wave of incident direction $\theta_q$. 

A straightforward beamforming approach would be to compute one array matrix for each frequency consisting of steering vectors for each direction. But in the eigenbeamforming approach the knowledge of the structure \eqref{eq:idealcoeffs} in the CH coefficients can be used to separate frequency-dependent from direction-dependent terms. Assuming for a moment the ideal case $\tilde{c}_p(k,R,\theta_q) = c_p(k,R,\theta_q)$, then the equalization
\begin{equation}
\label{eq:idealeq}
z_p(k,R,\theta_q) := \frac{1}{j^p J_p(kR)} c_p(k,R,\theta_q) = p_0 e^{-jp\theta_q}
\end{equation}
removes all frequency-dependent components and only leaves a purely direction-dependent term. For practical applications Parthy et al.~\cite{Parthy2011} propose
\begin{equation}
\label{eq:normalizedcoeffs}
\tilde{z}_p(k,R,\theta_q) := \frac{(-j)^p J_p(kR)}{\|J_p(kR)\|^2 + \beta}\; \tilde{c}_p(k,R,\theta_q), \quad \beta>0.
\end{equation}
with a smoothing factor $\beta > 0$. The need for this regularization lies in the fact that the Bessel function of higher orders and with them the denominator are close to zero for small values of $kR$, thus causing the equalization to become numerically unstable. Following the proposition of \cite{Torres2012}, we collect the $M$-dimensional microphone array input $\mathbf{y}_n$ at time instance $n$, apply an STFT and compute a vector of sampled CH coefficients $\mathbf{c}_n(k,R) \in \mathbb{C}^{2L+1}$ according to \eqref{eq:sampledcoeffs} by applying an FFT on $\mathbf{y}_n$. We then perform the equalization of \eqref{eq:normalizedcoeffs} to obtain $\mathbf{z}_n(k,R) \in \mathbb{C}^{2L+1}$. These preprocessing steps need to be done for each frequency, but the computational effort of the FFT operation is negligible compared to the subsequent sparse coding step. In order to estimate the incident direction from the equalized coefficients we now define the array matrix of steering vectors, which for the case of $q=1,\dots,Q$ considered angles and $p=-L,\dots,L$ observed modes writes as follows
\begin{equation}
D \in \mathbb{C}^{(2L+1) \times Q}, \quad \left[D\right]_{p,q} = e^{-jp\theta_q}.
\end{equation}
If we interpret this steering matrix as a dictionary, the sparse signal model in the CH domain is
\begin{equation}
\label{eq:CHL1SVDvec}
\mathbf{z}_n(k,R) = D \mathbf{s}_n(k,R) + \mathbf{w}_n,
\end{equation}
where $\mathbf{s}_n(k,R)$ is a sparse vector whose support indicates the position of the sources contained in the $n$-th frame at a certain wave number $k$ for an array radius of $R$, and $\mathbf{w}_n$ is additive Gaussian noise. At first sight this seems to be the same model as \eqref{eq:L1SVD}, only that the input data is now in the modal domain instead of the time-frequency domain. However, due to the coefficient equalization \eqref{eq:normalizedcoeffs} the steering matrix in \eqref{eq:CHL1SVDvec} is frequency-independent, i.e.~it remains unchanged for all frequencies. This allows us to concatenate the input vectors of all wave numbers $k=k_1,\dots,k_\Omega$ to one data matrix
\begin{equation}
Z_n := \begin{bmatrix} \mathbf{z}_n(k_1 R) & \dots & \mathbf{z}_n(k_\Omega R) \end{bmatrix} \in \mathbb{C}^{(2L+1) \times \Omega}
\end{equation}
and process them jointly within the same data model.

So far only data for a single time step has been considered. However, block-wise processing can increase the robustness of the DOA estimation if the observed signals are stationary over a certain observation period of $N$ frames. Thus, from now on we consider an $M \times N$-dimensional input matrix $Y = \begin{bmatrix}\mathbf{y} (t_1) & \dots & \mathbf{y} (t_N) \end{bmatrix}$, which in the CH domain leads to the cross-frequency data model
\begin{equation}
Z = DS
\end{equation}
with the $(2L+1) \times (\Omega N)$-dimensional CH coefficient matrix
\begin{equation}
Z := \begin{bmatrix} Z_1 & Z_2 & \dots & Z_N \end{bmatrix}.
\end{equation}
We assume that ideally, all columns of $S$ share the same support and the indices of the non-zero rows correspond to the positions of the respective sources in the mixture. In \cite{Malioutov2005} a method is proposed for reducing the computational complexity of estimating the support of $S$ by applying an SVD to the input data beforehand. The authors keep only the $K$-dimensional signal subspace, but report that overestimating $K$ up to $M-1$ does not deteriorate the results. Similarly, we decompose $Z=U_Z \Sigma_Z V_Z^\top$ and keep the square matrix $\tilde{Z} := U_Z \Sigma_Z$.
Thus we aim at minimizing $\|\tilde{Z}-DS\|_F$ under the condition that $S \in \mathbb{C}^{Q \times (2L+1)}$ is a row-sparse matrix. The convex relaxation of this problem is
\begin{equation}
\hat{S}=\min_S \|\tilde{Z} - D S\|^2 + \lambda \|S\|_{2,1}
\end{equation}
with $\|S\|_{2,1} = \sum_i \sqrt{\sum_j (s_{ij})^2}$ and a parameter $\lambda \in \mathbf{R}$ that weighs between the fidelity of the sparse approximation and the row-sparsity of $S$. After solving for $S$ the incident direction of the sound sources can be estimated from the support of
\begin{equation}
\hat{\mathbf{s}}:=\sum_{j=1}^{2L+1} |\mathbf{s}_j|. 
\end{equation}
Note that taking the sum over the columns of $|S|$ does not introduce further ambiguities in the result since the $\ell_{2,1}$-regularization enforces the columns to share the same support. In a nutshell, our Wideband Audio Sparse Circular Harmonics Localizer (WASCHL) solves a joint optimization problem for $M \times \Omega N$-dimensional time-frequency input data instead of computing the support and estimating the DOA separately at each frequency and for each input frame. Thereby, a frequency coherent DOA estimate is obtained at tremendously reduced computational effort.

\section{Discussion and Experiments}
Since the proposed solution of fast and coherent frequency-processing almost seems too good to be true it needs to be determined how far the discrete nature of a microphone array and the consequently incomplete sampling of the CH coefficients challenges the performance of our algorithm in comparison to the L1-SVD approach \cite{Malioutov2005}. On the other hand we compare our method to the circular harmonics beamformer (CHB, \cite{Torres2012}). While the response of a delay-and-sum (DSB) beamformer is rather widespread with significant sidelobes, eigenbeamforming using circular harmonics offers a narrower response with reduced sidelobes. Still, both methods share difficulties with distinguishing sources that are close together. Below a certain angular distance the response of the two sources merges into one common beam, such that they are detected as one. We refer to this as the angular distinction limit. In contrast to this, superresolution DOA methods such as MUSIC \cite{Schmidt1986}, ESPRIT \cite{Roy1989} or L1-SVD \cite{Malioutov2005}, manage to distinguish sources that are placed tightly together. Investigating for the cause, one finds that different frequencies have different influence on the localization result. At the lower end of the frequency range sources can be detected uniquely but rather coarse, whereas at higher frequencies the spatial resolution is increased, but aliasing occurs, i.e. the position estimates become ambiguous. Averaging over multiple frequencies can resolve these ambiguities, however at a much higher computational effort \cite{Malioutov2005} and at the risk of introducing additional errors. Our approach belongs to the class of super-resolution methods and therefore should exhibit a similarly high resolution. However, as discussed before, the finite number of microphones limits the order of modes that can be observed. At lower frequencies this does not introduce a significant error, since the Bessel functions have no significant influence. But at high frequencies the errors increase since the contributions of higher-order coefficients are no longer negligible and also there are zeros in the Bessel functions of lower orders. Thus we assume that the general angular resolution of our approach is about as good as the L1-SVD method, but there might be a significant angular distinction limit.
\subsection{Simulation}
In the following we consider a microphone array of $M=8$ microphones placed equidistantly in a circle of radius $R=0.12$\,m, which record signals at $16$\,kHz. The STFT is computed with a window length of $512$ samples at $50\%$ overlap, and the observation duration is $N=180$ frames (cf. \cite{Malioutov2005}) with $50\%$ overlap. $Q=360$ positions are evaluated at once, such that the minimum angular distance is one degree. Empirically, we choose the parameters $\beta=0.01$ (CHB, WASCHL) and $\lambda=1.1$ (L1-SVD, WASCHL).

\begin{figure}
        \centering
        \begin{subfigure}[b]{0.22\textwidth}
                \includegraphics[width=\columnwidth]{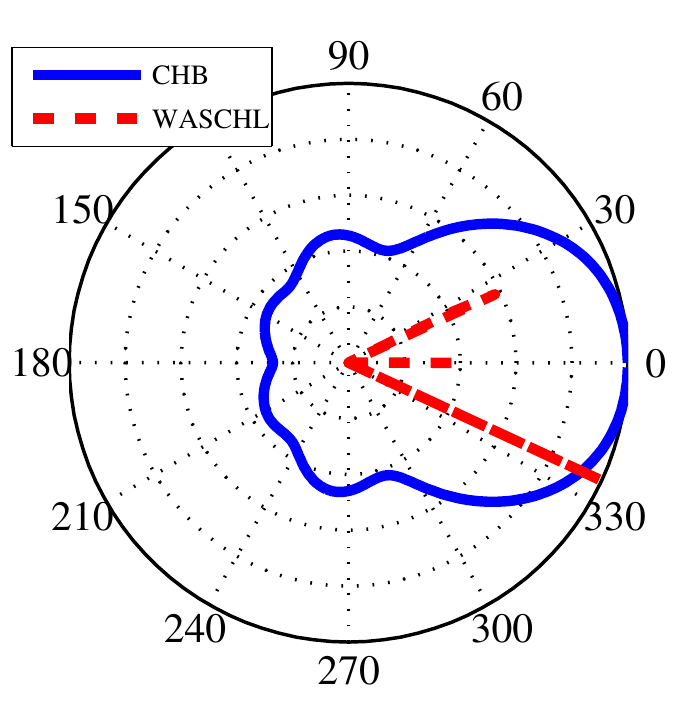}
                \caption{$\phi = 25^\circ$}
                \label{fig:resCHL1SVD}
        \end{subfigure}
        ~ 
        \begin{subfigure}[b]{0.22\textwidth}
                \includegraphics[width=\columnwidth]{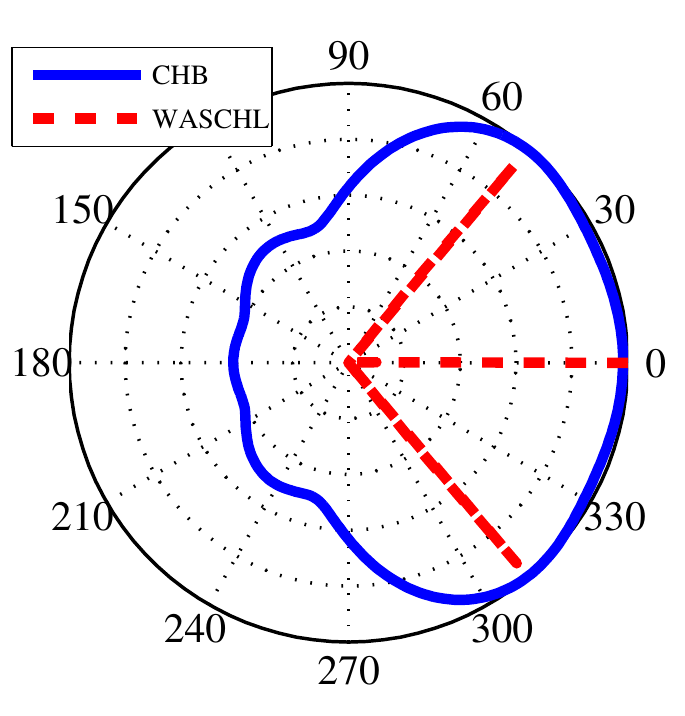}
                \caption{$\phi = 48^\circ$}
                \label{fig:resCHB}
        \end{subfigure}
        \caption{Simulated localization results for three sound sources at ($-\phi^\circ, 0^\circ, +\phi^\circ$) illustrating the minimum angular distinction limits for \WASCHL{} and CHB}
        \label{fig:res}
\end{figure}

Figure \ref{fig:res} shows the results of a simulation in which three sound sources (uncorrelated pink noise) are placed at $0^\circ$ and $\pm \phi$, respectively. This angular distance of the outer sources is varied up to the point where they can just be distinguished as separate sources by the respective localizer. As can be seen, the CHB resolves the sources up to about $\phi_{\textrm{CHB}}=45^\circ$. At an angular distance of $\phi=25^\circ$ the CHB merges all sources into one, while \WASCHL{} (and L1-SVD, not shown) resolve all three sources. For the case of $M=8$ microphones this is the angular distinction limit $\phi_{\textrm{\WASCHL{}},8}$ for our method, but using more microphones (e.g.~$M=24$) would help to improve it ($\phi_{\textrm{WASCHL},24}=10^\circ, \phi_{\textrm{CHB},24}=20^\circ$).
\subsection{Real recordings}
In a real-world scenario, i.e.~when continuous DOA estimation is performed, computational capacity is the limiting factor rather than the minimum angular resolution. For these purposes a tradeoff needs to be made between resolution, accuracy and processing time. In a first experiment we localize $24$ sequences of spoken human utterances in a common office room to compare the accuracy and the processing time of CHB, L1-SVD and our method \WASCHL{}. In each sequence, three speakers talk for about $5$ seconds from fixed positions ($45^\circ, 135^\circ, 225^\circ$), so that $6$ estimates of the three angles are obtained over time. We choose $K=7$, since we do not want to use a priori knowledge about the number of sources, and because overestimation does not hinder successful localization \cite{Malioutov2005}. For the final estimate we average over all frequencies (CHB and L1-SVD), and over the observation duration (CHB only), in order to allow for the fairest possible comparison. A standard peak finding routine extracts the three most dominant peaks that are at least $25^\circ$ apart for each observation. We assign the output to the closest permutation of ground truth angles and compute the absolute angular deviation. All algorithms are implemented\footnote{The MATLAB implementation of WASCHL and two demonstrations are available on the authors' web page \url{http://www.gol.ei.tum.de}} in MATLAB and we use the CVX toolbox \cite{cvx2013,Grant2008} to solve the second order cone program of the $\ell_{2,1}$-penalized cost function, cf.~\cite{Malioutov2005}. We measure the execution time of each algorithm on a desktop computer in order to compare the computational effort.

\begin{table}
\begin{center}
\begin{tabular}{|l|c|c|c|}
  \hline
  & CHB & \WASCHL{} & L1-SVD  \\
  \hline
  computation time & $1.5$s & $14$s & $2943$s \\
  \hline
  acc($\phi \leq 10^\circ$) & $94\%$ & $94\%$ & $99\%$\\
  \hline
  acc($\phi \leq 5^\circ$) & $76\%$ & $87\%$ & $98\%$\\
  \hline
  acc($\phi \leq 2^\circ$) & $41\%$ & $64\%$ & $95\%$\\
  \hline
\end{tabular}
\end{center}
\caption{Comparison of computation time and accuracy for localizing three human speakers in an office room}
\label{tab:threespeakers}
\end{table}

\begin{figure*}
        \centering
        \begin{subfigure}[b]{0.3\textwidth}
                \includegraphics[height=.63\textwidth]{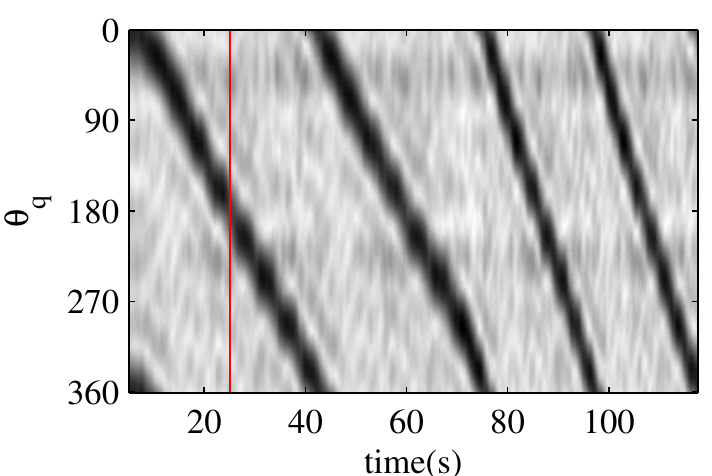}
				\caption{CHB}                
                \label{fig:ambCHB}
        \end{subfigure}%
        ~ 
        \begin{subfigure}[b]{0.3\textwidth}
                \includegraphics[height=.63\textwidth]{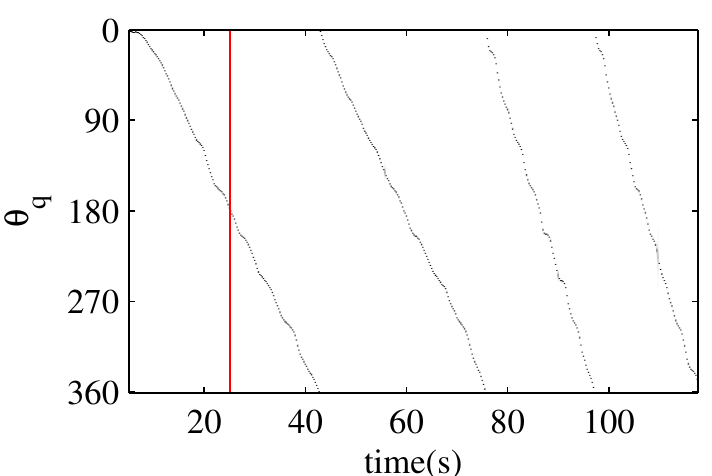}
			\caption{\WASCHL{}}                
                \label{fig:ambCHL1SVD}
        \end{subfigure}\quad \quad
        \begin{subfigure}[b]{0.3\textwidth}
                \includegraphics[height=.63\textwidth]{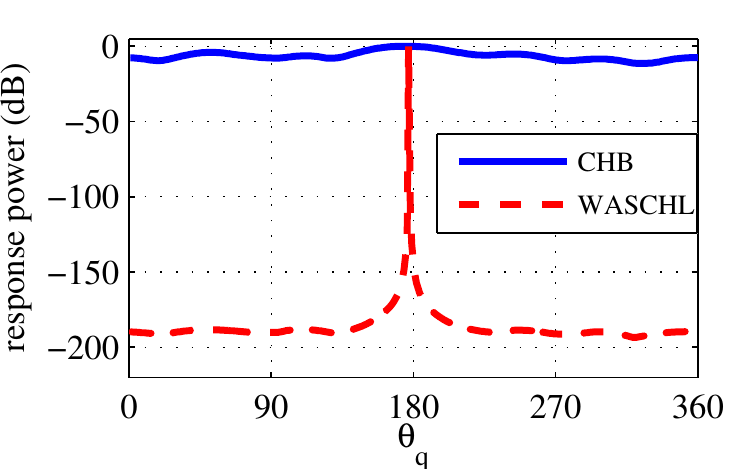}
				\caption{Response comparison at t=$24$\,s}                
                \label{fig:ambSlice}
        \end{subfigure}
        \caption{Responses for localizing a moving emergency vehicle with CHB and WASCHL. The vertical line in (a), (b) marks the observation time of (c)}\label{fig:amb}
\end{figure*}

Table~\ref{tab:threespeakers} shows the average processing times and the percentage of correct localization within certain bounds. A first conclusion that can be drawn is that the L1-SVD method clearly gives the most accurate estimation results, and as discussed before it has almost no limitation on the spatial placement of sources. However, its processing time is tremendous and thus prevents using it in a practical application. The CHB in contrast delivers extremely quickly, but the short computation time comes at the price of rather coarse angular resolution. Our method comes second both in accuracy and in processing time. But even if the accuracy seems to be half way between the competing methods it has to be noted that a spiky response allows for a much easier peak detection, which brings an additional advantage compared to the CHB, see Figures~\ref{fig:res} and \ref{fig:ambSlice}. Also, the processing time of our method is much closer to realtime performance, which should allow using it in a practical application if the implementation is optimized.

In a last experiment we compare our method against the CHB in a task of continuously localizing a moving sound source. A microphone array ($R=0.1$\,m, $M=8$) is placed outside on a parking lot while an emergency vehicle with activated alarm signal drives around it in circles at $20-30$ meters distance. The observation window is reduced to $30$~frames with an overlap of $5$~frames to account for the movement of the vehicle and to reduce the processing delay. Figures~\ref{fig:ambCHB} and \ref{fig:ambCHL1SVD} show the output of the CHB and our algorithm, respectively.
Although no ground truth position of the vehicle is available, it can be seen that the DOA estimates of both algorithms are in good agreement. Again, the localization result of \WASCHL{} shows a much narrower localization peak whereas the broad peak and several sidelobes can be seen in the CHB output. Finally, the momentary observation in Figure~\ref{fig:ambSlice} illustrates the spikiness of the WASCHL response compared to the CHB.

\section{Conclusion}
We propose the \WASCHL{} (Wideband Audio Sparse Circular Harmonics Localizer) method that combines the concept of wavefield analysis and sparsity-exploiting DOA estimation. The approach allows for frequency-coherent localization of an unknown number of sound sources using a circular microphone array. Due to the evaluation of circular harmonic coefficients instead of time delays all frequency components of a wideband audio signal can be processed at once. Experiments prove that the approach is computationally much cheaper than previous methods that process each frequency separately, while delivering much more precise estimates than low-complexity beamforming methods.

\end{document}